\documentclass{article}
\usepackage{spconf}

\usepackage[english]{babel}

\usepackage{ifpdf}

\usepackage{cite} 
\usepackage{url}
\usepackage{hyperref}

\usepackage[pdftex]{graphicx}
\graphicspath{{./figures/}}
\usepackage{color}
\usepackage{pgf, tikz, pgfplots}
\usetikzlibrary{shapes, arrows, automata, plotmarks}
\usetikzlibrary{calc,hobby,decorations}

\usepackage[cmex10]{amsmath}
\usepackage{amsfonts, amssymb, amsthm}
\usepackage{mathrsfs}

\usepackage{algorithm,algpseudocode}
\algnewcommand{\LeftComment}[1]{\Statex \(\triangleright\) #1}

\usepackage[margin=15mm]{geometry}
\usepackage{enumerate}
\usepackage{multirow}
\usepackage{rotating}
\usepackage{subcaption}
\input{auxFiles/pennColors.sty}
\usepackage{needspace}

\newcommand{\myparagraph}[1]{\needspace{1\baselineskip}\medskip\noindent {\bf #1}}

\input{auxFiles/mySymbol.sty}

\def\myfactor{1.0}
\def\unit{\myfactor cm}

\pgfplotsset{ ylabel near ticks,                 
              xlabel near ticks,                 
              tick label style = {font=\footnotesize},   
              label style = {font=\footnotesize}, 
              title style = {font=\footnotesize},
            }

\tikzstyle{empty node} = [ circle, 
                     draw = black,
                     text = black, 
                     minimum size = 0.8*\unit]

\tikzstyle{node} = [ empty node, 
                     fill = blue!50,
                     draw = blue!50,
                     text = white]

\tikzstyle{blue node} = [ empty node, 
                         fill = blue!50,
                         draw = blue!50,
                         text = white]

\tikzstyle{red node} = [ empty node, 
                         fill = red!50,
                         draw = red!50,
                         text = white]

\tikzstyle{green node} = [ empty node, 
                         fill = mygreen!50,
                         draw = mygreen!50,
                         text = white]

\tikzstyle{black node} = [ empty node, 
                           fill = black!40,
                           draw = black!40,
                           text = white]

\tikzstyle{empty dot} = [ circle, 
                           draw = black,
                           inner sep = 0pt,
                           anchor = center,
                           minimum size = 0.4*\unit]

\tikzstyle{dot} = [ empty dot, 
                    fill = blue!50,
                    draw = blue!50]

\tikzstyle{blue dot} = [ empty dot, 
                         fill = blue!50,
                         draw = blue!50]

\tikzstyle{red dot} = [ empty dot, 
                        fill = red!50,
                        draw = red!50]

\tikzstyle{black dot} = [ empty dot, 
                          fill = black!50,
                          draw = black!50]

\tikzstyle{edge}                 = [shorten >=1pt, shorten <=1pt]
\tikzstyle{directed edge}        = [edge, -stealth]
\tikzstyle{double directed edge} = [edge, stealth-stealth]
\tikzstyle{tight edge}                 = [shorten >=0pt, shorten <=0pt]

\tikzstyle{block} = [ rectangle,
                      minimum width = \unit,
                      minimum height = \unit,
                      fill = blue!15,
                      draw = black,
                      text = black]

\def\Tr{\mathsf{T}}

\def\Nul{\mathrm{Nul}}

\newtheorem{theorem}{\hspace{0pt}\bf Theorem}
\newtheorem{corollary}{\hspace{0pt}\bf Corollary}

\newtheorem{definition}{\hspace{0pt}\bf Definition}

\title{DISCRIMINABILITY OF SINGLE-LAYER GRAPH NEURAL NETWORKS}
\twoauthors
  {Samuel Pfrommer, Alejandro Ribeiro\thanks{Supported by NSF CCF 1717120, ARO W911NF1710438, ARL DCIST CRA W911NF-17-2-0181, ISTC-WAS and Intel DevCloud.}}
	{University of Pennsylvania\\
	Dept. of Electrical and Systems Engineering\\
	Philadelphia, PA}
  {Fer\hspace{0.015cm}nando Gama}
	{University of California, Berkeley\\
	Electrical Eng. and Computer Sci. Dept.\\
	Berkeley, CA}
\begin{document}
\ninept
\maketitle
\begin{abstract}
	Network data can be conveniently modeled as a graph signal, where data values are assigned to the nodes of a graph describing the underlying network topology. Successful learning from network data requires methods that effectively exploit this graph structure. Graph neural networks (GNNs) provide one such method and have exhibited promising performance on a wide range of problems. Understanding why GNNs work is of paramount importance, particularly in applications involving physical networks. We focus on the property of discriminability and establish conditions under which the inclusion of pointwise nonlinearities to a stable graph filter bank leads to an increased discriminative capacity for high-eigenvalue content. We define a notion of discriminability tied to the stability of the architecture, show that GNNs are at least as discriminative as linear graph filter banks, and characterize the signals that cannot be discriminated by either.
\end{abstract}
%


\section{Introduction} \label{sec:intro}



Data generated by networks are increasingly common in power grids \cite{Owerko20-Power}, robotics \cite{Tolstaya19-Flocking, Li20-Planning}, and wireless sensor networks \cite{Owerko18-Sensor, Eisen20-REGNN} among others \cite{Sandryhaila13-DSPG, Shuman13-SPG}. The irregular and complex nature of these data poses unique challenges that can only be addressed by incorporating underlying graph structure into the inner mechanisms of the machine learning model.

Graphs are used as a mathematical description of network topologies, while the data can be seen as a signal on top of the graph nodes. In a power grid \cite{Owerko20-Power}, for instance, generators and consumers can be modeled as nodes (buses), their electrical lines as edges, and the power generated (or consumed) as graph signals. Processing such data by accounting for the underlying network structure has been the goal of the field of \emph{graph signal processing} (GSP) \cite{Sandryhaila13-DSPG}. GSP extensions accounting for underlying topology have been developed for many familiar signal processing concepts, including the Fourier transform, convolutions, and filters \cite{Sandryhaila13-DSPG, Sandryhaila14-DSPGfreq}.

\emph{Graph convolutional} neural networks (GNNs) build upon graph convolutions to efficiently incorporate graph structure into the learning process \cite{Bronstein17-GeometricDeepLearning, Gama20-GNNs}. GNNs consist of a concatenation of layers, in which each layer applies a \emph{graph convolution} followed by a pointwise nonlinearity \cite{Bruna14-DeepSpectralNetworks, Defferrard17-ChebNets, Gama19-Archit, Kipf17-GCN}. GNNs have found promising performance in a myriad of applications including text categorization \cite{Defferrard17-ChebNets, Gama19-Archit}, clustering of citation networks \cite{Kipf17-GCN, Velickovic18-GAT, Vignac20-Choice}, authorship attribution \cite{Ruiz20-Nonlinear}, recommendation systems \cite{Ying18-LargeScaleGNN, Monti17-Movie} and source localization \cite{Isufi20-EdgeNets}.

The popularity of GNNs has been rooted in their observed success at a wide array of tasks involving graph data. Understanding the reasons why GNNs perform well on problems involving network data is thus a crucial research direction. This is especially true for applications that relate to physical systems and might therefore pose safety concerns if the limitations of GNNs are not appropriately understood. One aspect of understanding GNNs is determining the salient characteristics of the \emph{representation space} (i.e. the space of all possible functions or \emph{representation mappings} that can be learned by means of a GNN). For instance, we know that graph filters are naturally local and distributed information processing architectures, and since pointwise nonlinearities do not affect this property, GNNs inherit it \cite{Sandryhaila13-DSPG, Gama19-Archit}. This makes GNNs particularly attractive in scenarios where a decentralized solution is of essence \cite{Witsenhausen68-Counterexample}. We also know that GNNs are permutation equivariant \cite{ZouLerman19-Scattering, Gama20-Stability} which means that they exploit the internal symmetries of the underlying graph structure, and that they are stable to small changes in the graph support \cite{Gama20-Stability, Thanou20-Stability}. Additionally, GNNs are good at transferring to unseen graph topologies as long as they are similar to the ones observed during training \cite{Levie20-Transferability, Ruiz20-Transferability}. Another fundamental aspect of characterizing the representation space of GNNs is to understand the limitations of such representations. In this context, the conditions for identifying graph isomorphisms have been outlined \cite{Xu19-GIN, Hamilton19-Weisfeiler, Villar19-EquivIsomorphism} and graph scattering transforms have been used to explore the limits of the permutation equivariance and stability properties \cite{Wolf19-UnderstandingScattering}.

In this paper we focus on the property of discriminability. More specifically, we establish the conditions under which the inclusion of pointwise nonlinearities to a stable bank of graph filters (single-layer GNN) can lead to an increased discriminability of high-eigenvalue content. To do so, we first define a notion of discriminability tied to the degree of stability of the graph filter bank [cf. \eqref{eq:nondiscriminableSet}]. Then, we prove that GNNs are, at the very least, as discriminative as graph filter banks (Theorem~\ref{thm:firstStep}), we characterize the signals that the GNN will not be able to discriminate (Theorem~\ref{thm:secondStep}), we establish a case where the GNN is exactly as discriminative as the graph filter bank (Corollary~\ref{cor:setEquivalence}), and we show a practical case where the GNN is guaranteed to be more discriminative than the graph filter bank (Corollary~\ref{cor:tanhSet}).

Section~\ref{sec:GNN} provides a brief overview of GNNs, and in Section~\ref{sec:disciminability} we define discriminability and we state and analyze the main theoretical contributions (i.e. increased discriminability potential of GNNs). Section~\ref{sec:sims} describes numerical experiments that support our analysis. We conclude the paper in Section~\ref{sec:conclusions}. Proofs can be found in the Appendix.


\section{Graph Neural Networks} \label{sec:GNN}



Let $\ccalG = (\ccalV, \ccalE, \ccalW)$ be an undirected graph with $\ccalV = \{v_{1}, \ldots, v_{N}\}$ the set of $N$ nodes, $\ccalE \subseteq \ccalV \times \ccalV$ the set of edges, and $\ccalW: \ccalE \to \reals_{+}$ a function that assigns a positive weight to each edge. This graph represents the network structure on which the data is generated. The data itself is modeled as a graph signal $\bbx: \ccalV \to \reals$, which can be equivalently described as a vector $\bbx \in \reals^{N}$ such that its $i$th element $[\bbx]_{i} = \bbx(v_{i})$ is the value assigned to node $v_{i}$ \cite{Sandryhaila13-DSPG, Shuman13-SPG}.

The graph signal vector $\bbx$ alone contains no information regarding the underlying graph on which it is supported. To incorporate this structural information when processing $\bbx$, we define a \emph{support matrix} $\bbS \in \reals^{N \times N}$. This matrix respects the sparsity pattern of the graph, i.e. $[\bbS]_{ij} = 0$ whenever $(v_{j},v_{i}) \notin \ccalE$. In other words, the support matrix $\bbS$ is such that the only nonzero entries of the matrix are those $(i,j)$ entries corresponding to nodes such that $(v_{j},v_{i}) \in \ccalE$. This implies that $\bbS$ can be used to define a local and distributed linear operation as
\begin{equation} \label{eq:graphShift}
    [\bbS \bbx]_{i} = \sum_{j=1}^{N} [\bbS]_{ij} [\bbx]_{j} = \sum_{j : v_{j} \in \ccalN_{i}} [\bbS]_{ij} [\bbx]_{j}
\end{equation}
where $\ccalN_{i} = \{v_{j} \in \ccalV: (v_{j},v_{i}) \in \ccalE\} \cup \{v_{i}\}$ is the neighborhood of node $v_{i}$. Note that, due to the sparsity pattern of $\bbS$, matrix multiplication in \eqref{eq:graphShift} only involves a linear combination of signal values in neighboring nodes. Thus, the linear operation \eqref{eq:graphShift} is local, because it only requires information from $\ccalN_{i}$, and is distributed, since the output $[\bbS \bbx]_{i}$ can be computed separately at each node $v_{i}$. In the graph signal processing (GSP) literature, the operation \eqref{eq:graphShift} is known as a \emph{graph shift} (because it generalizes the elementary time shift upon which traditional signal processing is built), and thus the matrix $\bbS$ often receives the name of \emph{graph shift operator} (GSO) \cite{Sandryhaila13-DSPG, Shuman13-SPG}. Typical choices of support matrix include the adjacency \cite{Sandryhaila13-DSPG} and the graph Laplacian \cite{Shuman13-SPG}, as well as their normalized counterparts \cite{Defferrard17-ChebNets, Kipf17-GCN}.

In analogy to traditional signal processing, we leverage operation \eqref{eq:graphShift} to define a finite impulse response (FIR) graph filter as a linear combination of shifted versions of the signal \cite{Isufi17-ARMA}
\begin{equation} \label{eq:graphFilter}
    \bbH(\bbS) = \sum_{k=0}^{K} h_{k} \bbS^{k}.
\end{equation}
The FIR graph filter \eqref{eq:graphFilter} defines a linear mapping $\bby = \bbH(\bbS) \bbx$ between graph signals. Note that $\bbS^{k}\bbx = \bbS (\bbS^{k-1} \bbx)$ so that the output $[\bbS^{k}\bbx]_{i}$ at node $v_{i}$ can be computed by means of $k$ repeated exchanges with one-hop neighbors in $\ccalN_{i}$ [cf. \eqref{eq:graphShift}]. This makes the FIR graph filter a local and distributed linear operation as well. Even though there are other non-FIR graph filters \cite{Coutino19-Distributed, Isufi20-EdgeNets}, from now on we refer to \eqref{eq:graphFilter} as simply a graph filter or a \emph{graph convolution} \cite{Gama20-GNNs}.

Oftentimes graph signals can be better analyzed in the \emph{graph frequency domain} \cite{Sandryhaila14-DSPGfreq}. To do this, let $\bbS = \bbV \bbLambda \bbV^{\Tr}$ be the eigendecomposition of the support matrix, where $\bbV = [\bbv_{1},\ldots,\bbv_{N}]$ contains the eigenvectors $\bbv_{i}$ such that $\bbS \bbv_{i}=\lambda_{i}\bbv_{i}$ for $\lambda_{i}$ representing the $i$th eigenvalue contained in the entries of the diagonal matrix $\bbLambda$. We consider the eigenvalues to be ordered by $|\lambda_{1}| \leq \cdots \leq |\lambda_{N}|$. The \emph{graph Fourier transform} (GFT) of a graph signal is given by its projection onto the eigenbasis, i.e. $\tbx = \bbV^{\Tr} \bbx$. The GFT of the output of a graph filter then becomes \cite{Sandryhaila14-DSPGfreq}
\begin{equation} \label{eq:GFToutput}
    \tby 
        = \bbV^{\Tr} \bby 
        = \bbV^{\Tr} \bbH(\bbS) \bbx 
        = \sum_{k=0}^{K} h_k \bbLambda^{k} \tbx 
        = \bbH(\bbLambda) \tbx.
\end{equation}
Due to the diagonal nature of $\bbLambda$, the output GFT in \eqref{eq:GFToutput} can be computed as an entrywise product with the corresponding input GFT
\begin{equation} \label{eq:GFTpointwise}
    [\tby]_{i} 
        = \sum_{k=0}^{K} h_{k} \lambda_{i}^{k} [\tbx]_{i} 
        = h(\lambda_{i}) [\tbx]_{i}
\end{equation}
where $h: \reals \to \reals$ defines the \emph{frequency response} of the filter
\begin{equation} \label{eq:freqResponse}
    h(\lambda) = \sum_{k=0}^{K} h_{k} \lambda^{k}.
\end{equation}
Thus, we see from \eqref{eq:GFTpointwise} that the $i$th frequency component of the output $[\tby]_{i}$ can be computed by multiplication of the $i$th frequency component of the input $[\tbx]_{i}$ with the frequency response \eqref{eq:freqResponse} evaluated at the $i$th eigenvalue $h(\lambda_{i})$. Note that while the GFT of the signal depends on the eigenvectors of $\bbS$, the frequency response \eqref{eq:freqResponse} gets evaluated on the eigenvalues of $\bbS$. Moreover, the frequency response \eqref{eq:freqResponse} is defined entirely by the filter taps $\{h_{k}\}$, and the effect of the graph is observed by the particular eigenvalues on which $h$ is instantiated.

Graph filters are linear mappings between graph signals, and as such, their representation space can only capture linear dependencies between input and output. Graph neural networks (GNNs) \cite{Bruna14-DeepSpectralNetworks, Defferrard17-ChebNets, Gama19-Archit} cascade graph filters with pointwise nonlinearities as a means of constructing nonlinear mappings between input and output. Let $\sigma: \reals \to \reals$ be a nonlinear function, and $\{\bbH_{\ell}^{fg}(\bbS)\}_{f,g}$ be a bank of $F_{\ell-1} \times F_{\ell}$ filters, indexed by $\ell$. A GNN is formally defined as a cascade of $L$ blocks or \emph{layers} each of which applies a bank of filters, followed by a pointwise nonlinearity
\begin{equation} \label{eq:GNN}
    \bbx_{\ell}^{g} = \sigma \Big( \sum_{f=1}^{F_{\ell-1}} \bbH^{fg}(\bbS) \bbx_{\ell-1}^{f} \Big) \ , \ g = 1,\ldots, F_{\ell}
\end{equation}
for $\ell = 1,\ldots,L$, and where $[\sigma(\bbx)]_{i}=\sigma([\bbx]_{i})$ in a convenient abuse of notation. The input is $\bbx_{0} = \bbx$ with $F_{0} = 1$ and the output is collected at the output of the last layer, $\bbPhi(\bbx;\bbS) = \{\bbx_{L}^{g}\}_{g=1}^{F_{L}}$, creating a nonlinear map $\bbPhi: \reals^{N} \to \reals^{N \times F_{L}}$. Each signal $\bbx_{\ell}^{g}$ at the output of each layer is typically called a \emph{feature}. The specific nonlinear function $\sigma$, the number of layers $L$, the number of features $F_{\ell}$ per layer, and the number of filter taps $K_{\ell}$ per layer [cf. \eqref{eq:graphFilter}] are set by design.

GNNs exhibit several key properties that draw insights on their observed superior performance when processing graph signals. First, they inherit the local and distributed nature of graph filters [cf. \eqref{eq:graphFilter}], and thus its output can be computed separately at each node, relying only on repeated communication exchanges with one-hop neighbors. They are also permutation equivariant \cite{Xu19-GIN, Villar19-EquivIsomorphism, Gama20-Stability} meaning that they exploit the internal symmetries of the graph topological structure to improve learning. Furthermore, if the graph filters in the corresponding filter bank are integral Lipschitz (see Def.~\ref{def:ILfilters}), then the GNN is stable under perturbations of the graph support, meaning that the change in the output of the GNN caused by a change in the underlying graph support is linearly bounded by the size of the support change.

\begin{definition} \label{def:ILfilters}
    A graph filter $\bbH(\bbS)$ \eqref{eq:graphFilter} with frequency response $h(\lambda)$ \eqref{eq:freqResponse} is an integral Lipschitz filter if it satisfies
    \begin{equation} \label{eq:ILcondition}
    |h(\lambda_{2}) - h(\lambda_{1})| \leq C\frac{|\lambda_{2}-\lambda_{1}|}{|\lambda_{2}+\lambda_{1}|/2}
    \end{equation}
    for all $\lambda_{1},\lambda_{2} \in \reals$, and for some \emph{integral Lipschitz constant} $C > 0$.
\end{definition}

\noindent In short, integral Lipschitz filters are those whose frequency response is Lipschitz with a constant that depends on the midpoint of the interval $[\lambda_{1},\lambda_{2}]$. Certainly, the bigger the values of $\lambda$, the smaller the Lipschitz constant is, and thus the less variability the frequency response admits. Note that condition \eqref{eq:ILcondition} implies that $|\lambda h'(\lambda)| \leq C$, which alternatively shows that if $\lambda$ is large, then $h'(\lambda)$ has to be small for the product to be bounded or, similarly, that if $\lambda$ is small, then $h'(\lambda)$ can be large.



\section{Discriminability} \label{sec:disciminability}



Discriminability is concerned with the ability of a given processing architecture to tell two distinct signals apart. In particular, we are interested in comparing the discriminability of a bank of linear graph filters
\begin{equation} \label{eq:graphFilterBank}
    \ccalH(\bbx; \bbS) = \Big\{ \bbH^{f}(\bbS) \bbx \Big\}_{f=1}^{F}
\end{equation}
with the discriminability of a single-layer GNN [cf. \eqref{eq:GNN}]
\begin{equation} \label{eq:GNNsingle}
    \bbPhi(\bbx; \bbS) = \Big\{\sigma \big( \bbH^{f}(\bbS) \bbx \big) \Big\}_{f=1}^{F}.
\end{equation}
Both the graph filter bank $\ccalH$ \eqref{eq:graphFilterBank} and the GNN $\bbPhi$ \eqref{eq:GNNsingle} are stable under perturbations of the graph support $\bbS$ as long as the filters $\bbH^{f}(\bbS)$ are integral Lipschitz [cf. Def.~\ref{def:ILfilters}]. In what follows, we introduce a novel notion of discriminability as it pertains to the stability of the architecture (Sec.~\ref{subsec:discriminabilityDef}) and then we prove that GNNs are potentially more discriminable than graph filters (Sec.~\ref{subsec:discriminabilityThm}). We assume that there exists a readout layer that properly extracts information from the resulting features.


\subsection{Notion of discriminability} \label{subsec:discriminabilityDef}

\begin{figure}
    \centering
    \def \thisplotscale {3.68}

\def \unit {\thisplotscale cm}

\def \frequencyresponse 
{ 0.9*exp(-(6.0*(x-0.0))^2) }
\def \frequencyresponseTwo
{ 0.9*exp(-(3.0*(x-0.6))^2) }
\def \frequencyresponseThree
{ 0.9*exp(-(1.5*(x-1.8))^2) }
\def \frequencyresponseFour
{ 0.9*exp(-(1.0*(x-3.6))^2) }
\def \frequencyresponseFive
{ 0.9 - 0.9*exp(-(1.1*(x-1.9))^2) }

\begin{tikzpicture}[x = 1*\unit, y=1*\unit]
\begin{axis}[scale only axis,
             width  = 2.4*\unit,
             height = 0.8*\unit,
             xmin = -0.5, xmax=6.8,
             xtick = {0, 0.6, 1.8, 3.25, 3.6, 5.6},
             xticklabels = {$\lam_1=0\ $, $\lam_2$, $\lam_3$, {\color{pennred}$ \lambda_{C}$}, $\lam_4$, $\lam_5$},
             ymin = -0, ymax = 1.15,
             ytick = {-1}]

\addplot[domain=-0.5:0.6,samples = 80, color = pennblue, thick] {\frequencyresponse};
\addplot[domain=-0.5:1.8,samples = 80, color = pennblue, thick] {\frequencyresponseTwo};
\addplot[domain=-0.5:6.8,samples = 80, color = pennblue, thick] {\frequencyresponseThree};
\addplot[domain= 1.85:6.8,samples = 80, color = pennblue, thick] {\frequencyresponseFive};
\addplot[thick, samples=80, smooth, domain=0:6, color = pennred, dashed] coordinates {(3.25,-1)(3.25,3)};
\end{axis}
\end{tikzpicture}
    \caption{Bank of integral Lipschitz filters.}
    \label{fig:ILfilters}
\end{figure}
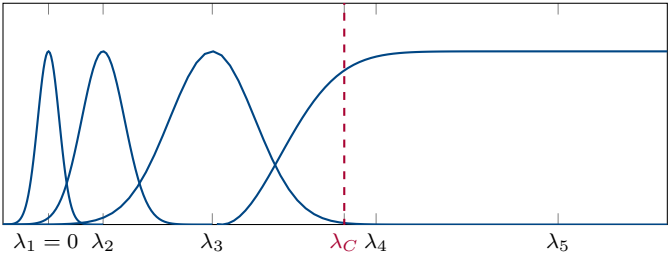

The stability of the graph filter bank $\ccalH$ and the GNN $\bbPhi$ are proportional to the integral Lipschitz constant $C$ such that a smaller value of $C$ leads to a more stable GNN, see \cite[Thms. 2, 4]{Gama20-Stability}. Thus, stable architecture have a low value of $C$. However, a small $C$ causes the filters to become flat for smaller values of $\lambda$, see Fig.~\ref{fig:ILfilters}. More specifically, as $\lambda$ increases, the derivative of the filter $h'(\lambda)$ has to decrease, since $|\lambda h'(\lambda)| \leq C$. If $h'(\lambda)$ is small, then the filter is nearly flat. We can then define the \emph{cutoff frequency} $\lambda_{C}(\varepsilon)$ to be the eigenvalue such that $h'(\lambda) < \varepsilon$ for all $\lambda > \lambda_{C}(\varepsilon)$, for some given $\varepsilon$. Certainly, the cutoff frequency is parametrized by the integral Lipschitz constant $C$ in a way that, for a fixed value of $\varepsilon$, we have $\lambda_{C'}(\varepsilon) < \lambda_{C}(\varepsilon)$ for $C' < C$. Therefore, the more stable the GNN is, the lower $C$ is, and the lower the cutoff frequency $\lambda_{C}$ is.

A small cutoff frequency implies that the filter bank is not able to tell apart signals whose difference lies in frequencies located above $\lambda_{C}$. This exhibits a tradeoff between stability (the need for a smaller $C$) and discriminability (the need for a higher $\lambda_{C}$) for integral Lipschitz filters. Let us fix $C$ and $\varepsilon$ such that, for the given support matrix $\bbS$, the eigenvalues satisfy
\begin{equation} \label{eq:ILconditionGSO}
|\lambda_{1}| \leq \cdots \leq |\lambda_{K}| < \lambda_{C}(\varepsilon) < |\lambda_{K+1}| \leq \cdots |\lambda_{N}|.
\end{equation}
This means that signals whose difference has frequency content located in eigenvalues greater than $\lambda_{C}$ cannot be discriminated, see Fig.~\ref{fig:ILfilters}. These are signals whose difference can be written as a linear combination of eigenvectors $\bbv_{K+1},\ldots,\bbv_{N}$, or equivalently, signals whose difference lies in the column space of $\bbV_{N-K} = [\bbv_{K+1},\ldots,\bbv_{N}]$. Noting that the column space of $\bbV_{N-K}$ is equivalent to the null space of $\bbV_{K} = [\bbv_{1},\ldots,\bbv_{K}]$, i.e., the eigenvectors associated to the discriminable frequencies, we can then define the \emph{set of nondiscriminable signals} as
\begin{equation} \label{eq:nondiscriminableSet}
\ccalD = \big\{ \bbx, \bby \in \reals^{N} : (\bbx - \bby) \in \Nul(\bbV_{K}) \big\}.
\end{equation}
Since a linear operator does not create frequency content, it is immediate that the set of nondiscriminable signals is unchanged by the use of a graph filter bank \eqref{eq:graphFilterBank}, i.e. $\ccalD \equiv \ccalD_{\ccalH}$ with
\begin{equation}  \label{eq:nondiscriminableSetFilter}
\ccalD_{\ccalH} = \Big\{ \bbx, \bby \in \reals^{N} : \big( \ccalH(\bbx;\bbS) - \ccalH(\bby;\bbS) \big) \in \Nul(\bbV_{K}) \Big\}.
\end{equation}
When considering GNNs of the form \eqref{eq:GNNsingle}, the set of nondiscriminable signals becomes
\begin{equation}  \label{eq:nondiscriminableSetGNN}
\ccalD_{\bbPhi} = \Big\{ \bbx, \bby \in \reals^{N} : \big(\bbPhi(\bbx;\bbS)  - \bbPhi(\bby;\bbS)\big) \in \Nul(\bbV_{K}) \Big\}.
\end{equation}
Due to the effect of the nonlinearity $\sigma$, the set $\ccalD_{\bbPhi}$ may be different from the set $\ccalD_{\ccalH}$ for the same filter bank $\{\bbH^{f}\}_{f=1}^{F}$.


\subsection{Enhanced discriminability of GNNs} \label{subsec:discriminabilityThm}

We now analyze and compare the discriminability of graph filters with that of GNNs. In particular, we first prove that any pair of signals that can be discriminated by the graph filter bank \eqref{eq:graphFilterBank} can also be discriminated by the GNN \eqref{eq:GNN}. Then, we characterize the signal pairs that cannot be discriminated by either the graph filter bank or the GNN. Finally, we prove that, for a $\tanh$ nonlinearity, the GNN is more discriminative than the graph filter bank.

We start by proving that GNNs are at least as discriminative as graph filters; i.e. there is no discriminabililty lost in adding a nonlinearity.
\begin{theorem} \label{thm:firstStep}
    Let $\{\bbH^{f}\}_{f=1}^{F}$ be a bank of $F$ integral Lipschitz graph filters \eqref{eq:graphFilterBank} with a constant $C$ such that \eqref{eq:ILconditionGSO} holds for the given support matrix $\bbS$. Let $\bbPhi$ be a one-layer GNN as in \eqref{eq:GNNsingle} with a Lipschitz continuous, strictly monotone nonlinearity $\sigma$. If at least one filter $\bbH^{f}$ has a frequency response such that $h^{f}(\lambda)=0$ for $\lambda> \lambda_{C}$, then it holds that
    \begin{equation} \label{eq:firstStep}
        (\bbx,\bby) \notin \ccalD_{\ccalH} \Rightarrow (\bbx,\bby) \notin \ccalD_{\bbPhi}
    \end{equation}
    for all $(\bbx,\bby) \notin \ccalD_{\ccalH}$, with $\ccalD_{\ccalH}$ and $\ccalD_{\bbPhi}$ defined as in \eqref{eq:nondiscriminableSetFilter} and \eqref{eq:nondiscriminableSetGNN}, respectively.
\end{theorem}
\begin{proof}
    See Appendix.
\end{proof}

\noindent Theorem~\ref{thm:firstStep} states that all pairs of signals that can be discriminated by the filter bank can also be discriminated by the GNN. More specifically, it suffices for the GNN to have only $F=1$ filter with $h^{1}(\lambda_{j}) =0$ for $j>K$ to be at least as discriminative as the corresponding linear filter bank. This is a sensible condition, since setting $h^{f}(\lambda) = 0$ for $\lambda> \lambda_{C}$ for at least one filter guarantees that no nonlinearity-generated low-eigenvalue content (generated from the high-eigenvalue content) interferes with the low-eigenvalue content (which can already be discriminated). In essence, Theorem~\ref{thm:firstStep} guarantees that the GNN is at least as discriminative as the graph filter bank, meaning that the nonlinearity does not decrease the discriminatory power.

Next, we characterize the pairs of signals that are not discriminable by either the graph filter bank or the GNN. 
\begin{theorem} \label{thm:secondStep}
    Let $\{\bbH^{f}\}_{f=1}^{F}$ be a bank of $F \geq 2$ integral Lipschitz graph filters \eqref{eq:graphFilterBank} with a constant $C$ such that \eqref{eq:ILconditionGSO} holds for the given support matrix $\bbS$. Assume the first filter satisfies $h^{1}(\lambda) = 0$ for $\lambda > \lambda_{C}$. Let $\bbPhi$ be a one-layer GNN as in \eqref{eq:GNNsingle} with a Lipschitz continuous, strictly monotone nonlinearity $\sigma$. Let $(\bbx,\bby) \in \ccalD_{\ccalH}$. Then,
    \begin{equation}  \label{eq:NDpairGNN}
        (\bbx,\bby) \in \ccalD_{\bbPhi} \quad \Leftrightarrow \quad b_{i}^{f} = b^{f} \ \forall\ i \in \{1,\ldots,N\}
    \end{equation}
    for all $f$ such that $h^{f}(\lambda) \neq 0$ for $\lambda > \lambda_{C}$, and where $b_{i}^{f} = (\sigma(x_{i}^{f}) - \sigma(y_{i}^{f}))/(x_{i}^{f}-y_{i}^{f})$ is the secant for $x_{i}^{f} - y_{i}^{f} \neq 0$ and $b_{i}^{f} = \sigma'(x_{i}^{f})$ is the derivative for $x_{i}^{f}=y_{i}^{f}$, with $x_{i}^{f} = [\bbH^{f}\bbx]_{i}$ and $y_{i}^{f} = [\bbH^{f}\bby]_{i}$, respectively.
\end{theorem}
\begin{proof}
    See Appendix.
\end{proof}
\noindent Theorem~\ref{thm:secondStep} plays a key role in characterizing the signals that we will not be able to discriminate, even when using a GNN. More specifically, Theorem~\ref{thm:secondStep} states that if the secant (or the derivative) of the nonlinearity, evaluated at the output of the graph filter, is the same at all nodes, then the pair of signals will not be discriminated. Note that since the value of $b_{i}^{f}$ depends on the value of the secant at the output of the filter $h^{f}$, having a larger graph filter bank (large $F$) with nonzero frequency responses beyond the cutoff frequency increases the possibility that at least one filter will have a distinct $b_{i}^{f}$.

In terms of filter training (and design), we need to guarantee that at least one filter in the bank has a nonzero frequency response beyond the cutoff frequency. Otherwise, all pair of signals that are nondiscriminable with a filter bank will also be nondiscriminable with the GNN.
\begin{corollary} \label{cor:setEquivalence}
    Under the setting of Theorem~\ref{thm:secondStep}, assume that all filters in the bank are such that $h^{f}(\lambda)=0$ for $\lambda > \lambda_{C}$, for all $f \in \{1,\ldots,F\}$. Then,
    \begin{equation} \label{eq:setEquivalence}
        \ccalD_{\ccalH} \equiv \ccalD_{\bbPhi}
    \end{equation}
\end{corollary}
\begin{proof}
    See Appendix.
\end{proof}
\noindent Corollary \eqref{cor:setEquivalence} states that having at least one filter with a nonzero frequency response beyond the cutoff frequency is a necessary condition for having a potentially more discriminative architecture. This makes sense, since if $h^{f}(\lambda) = 0$ for all $\lambda > \lambda_{C}$ and for all filters in the bank, then there is no filter that can actually pick up the high eigenvalue differences between $\bbx$ and $\bby$ (which are the only differences present, since $(\bbx,\bby)$ is in $\ccalD_{\ccalH}$ by hypothesis).

\begin{corollary} \label{cor:tanhSet}
    Under the setting of Theorem~\ref{thm:secondStep}, let $\sigma = \tanh$, and assume $\lambda_{C}(\varepsilon)$ and $\bbS$ are such that $N-K > 1$ [cf. \eqref{eq:ILconditionGSO}]. Then,
    \begin{equation} \label{eq:tanhSet}
        \ccalD_{\bbPhi} \subset \ccalD_{\ccalH}
    \end{equation}
\end{corollary}
\begin{proof}
    See Appendix.
\end{proof}
\noindent Corollary~\eqref{cor:tanhSet} shows one case in which the GNN is certifiably more discriminative than the graph filter.

In summary, we (i) proved that GNNs are, at the very least, as discriminative as graph filter banks (Theorem~\ref{thm:firstStep}), (ii) characterized the signals that the GNN will not be able to discriminate (Theorem~\ref{thm:secondStep}), (iii) established a case where the GNN is exactly as discriminative as the graph filter bank (Corollary~\ref{cor:setEquivalence}), (iv) shown a practical case where the GNN is guaranteed to be more discriminative than the graph filter bank (Corollary~\ref{cor:tanhSet}).


\section{Numerical Experiments} \label{sec:sims}



We support our theorems numerically by considering a synthetic regression problem where we control the subspace of generated signals. More precisely, we assume access to input-output pairs $\ccalT = \{(\bbx_{i},\bby_{i})\}$ that are related by some unknown function
\begin{equation}
    \bby = f (\bbx)
\end{equation}
The objective is to approximate the function $f$ in a distributed manner by employing either graph filters \eqref{eq:graphFilterBank} or GNNs \eqref{eq:GNNsingle}.

\myparagraph{Graph topology.} The underlying random graph is formed by distributing $N$ nodes uniformly over the unit square $[0,1]^2$ with each node adjacent to its $5$ nearest neighbors, producing a geometric (planar) random graph. Edge weights are assigned to be inversely proportional to the exponential of their separating distance. We consider the graph Laplacian $\bbL = \bbD - \bbA$, where $\bbD$ is the degree matrix and $\bbA$ is the weighted adjacency matrix. We normalize $\bbL$ by its largest eigenvalue, and adopt it as the graph shift operator $\bbS$ so that $\| \bbS \| = 1$.

\myparagraph{Signal generation.} In order to illustrate the theoretical results presented in this paper, we need to ensure that the $\bbx_{i}$ in $\{ (\bbx_i, \bby_i) \}$  is controlled to belong to a certain subspace. We thus construct $\bbx$ first, generating it in a particular subspace of interest and then applying an explicit model $f$ to obtain $\bby$. To isolate the high-eigenvalue subspace, we apply the projection operator $\bbV_{N-K} \bbV_{N-K}^{\Tr}$ onto random vectors $\bbw$ drawn from a Gaussian distribution with zero mean and unit variance
\begin{equation}
    \bbx = \frac{\bbV_{N-K} \bbV_{N-K}^{\Tr} \bbw}
                {\| \bbV_{N-K} \bbV_{N-K}^{\Tr} \bbw \|}
\end{equation}
where we normalize to control for scaling effects when $K \ll N$. Generating low-eigenvalue signals differs only by replacing $\bbV_{N-K}$ with $\bbV_K$. Finally, we let our target signal be given by
\begin{equation}
    \bby = f(\bbx) = sign( c_0 \bbI \bbx + c_1 \bbL \bbx + c_2 \bbL^2 \bbx ).
\end{equation}
Coefficients $c_i$ are drawn independently from a uniform distribution $[-1,1]$. We use the $sign$ function to add a nonlinear component that is distinct from our GNN activation function. The resulting target output graph signal is $\bby \in \{-1,1\}^{N}$.

\myparagraph{Models and training.} Our two models are given by \eqref{eq:graphFilterBank} and \eqref{eq:GNNsingle} with $F=32$ features, $3$ filter taps, and $\sigma = \tanh$ for the GNN. Both models combine their features with a single-tap readout layer and were trained using Adam \cite{Kingma15-ADAM} with a learning rate of $10^{-3}$ and per-epoch decay of $0.9$. The training dataset consisted of $8000$ samples generated by randomly sampling $\bbw$ as discussed, with $200$ samples used for validation and $200$ for testing. The experiment was repeated with 30 random graphs for each signal subspace, with $N=50$ and $K=10$; i.e., the upper quintile of eigenvalues were considered for our cutoff point $\lam_C$. Stability to perturbations was enforced by adding a regularizer proportional to the filter bank integral Lipschitz constant \eqref{eq:ILcondition} with weight $0.01$.

\begin{figure}
    \centering
    \includegraphics[width=0.5\textwidth]{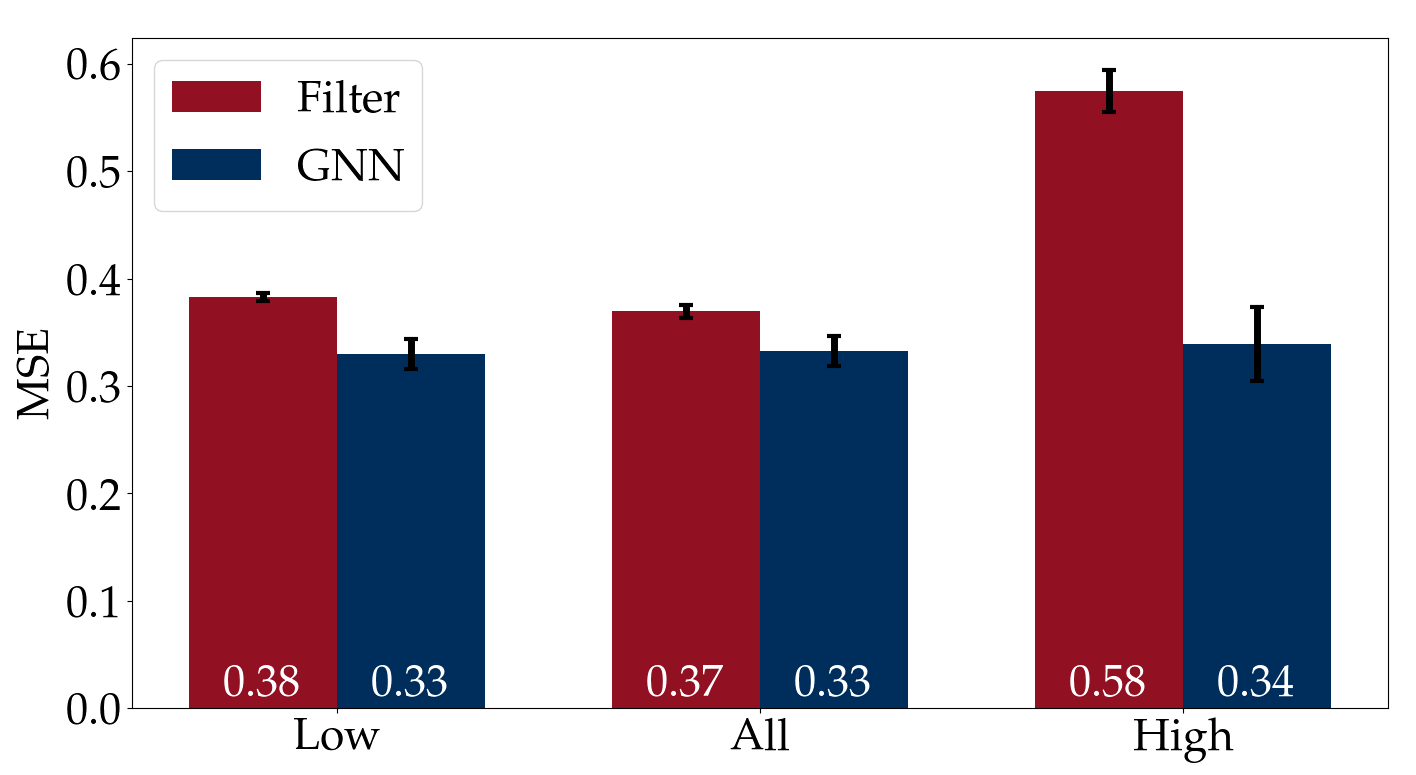}
    \caption{Test set errors are reported for input signals generated in different subspaces of the GSO's eigenspace. Vertical bars represent 95\% confidence intervals over 30 random geometric graph instantiations. When regularized with the integral Lipschitz penalty, linear graph filters perform noticeably worse than GNNs for high-eigenvalue signals. }
    \label{fig:results}
\end{figure}

\myparagraph{Discussion.} Results are summarized in Fig.~\ref{fig:results}. For a single-layer network, the continuous $\tanh$ nonlinearity provided marginal added benefit for estimating the discrete $sign$ function, leading to comparable performance for both low eigenvalues and the entire spectrum. However, for signals generated with high-eigenvalue content, the GNN error remains virtually unchanged, while the linear graph filter error increases by $57\%$. This empirically supports our observation that the introduction of nonlinearities provides GNNs with the ability to process high-eigenvalue content that is not discriminable with a regular graph filter.

Furthermore, we can infer that the secant condition necessary for increased discriminability in Theorem~\ref{thm:secondStep} does not restrict the practical ability of the GNN to discriminate signals. Namely, condition \eqref{eq:NDpairGNN} is generally satisfied. This matches expectation since intuitively it is highly improbable that a pair of input signals have precisely the same secants along all coordinates, and along all filters. Similarly, although the existence of a filter with zero response for $\lam > \lam_C$ is sufficient for GNNs to be at least as discriminative as graph filters by Theorem~\ref{thm:firstStep}, it is not necessary to explicitly enforce this condition. These observations indicate that nonlinearities are both practically and theoretically effective for producing architectures that are both stable to graph support perturbations and capable of discriminating high-eigenvalue content.

We emphasize that this is an illustrative, synthetic example whose main purpose is to showcase the implications of the theoretical concepts developed in the rest of the paper. The choice of a synthetic example allows us to control exactly the subspace on which the signals lie, and as such, to specifically illustrate the impact of low- and high-eigenvalue content: namely, that linear graph filters struggle to process signals with high-eigenvalue content whereas GNNs are successful.


\section{Conclusions} \label{sec:conclusions}



In this paper, we focused on analyzing the discriminability of graph filter banks and GNNs. We defined a notion of discriminability that is tied to the stability of the architecture. We proved that GNNs are, at the very least, as discriminative as graph filter banks and characterized the signals that the GNN will not be able to discriminate. We also characterized the uncommon case where the GNN is exactly as discriminative as the graph filter bank and established a practical case where the GNN is certified to be more discriminative than the graph filter bank.

This preliminary investigation opens up several exciting areas of future research. Namely, analyzing the existence of filters that can guarantee that the secants are always different at all nodes, characterizing the relationship between the number of filters in the bank and discriminability, and understanding the impact of specific nonlinearities (for instance, if there are nonlinearities that have a constant secant irrespective of the filter).


\bibliographystyle{bibFiles/IEEEbib}
\bibliography{bibFiles/myIEEEabrv,bibFiles/bibDiscriminability}


\clearpage
\newpage

\appendix

\section*{Appendix}



\begin{proof}[Proof of Theorem~\ref{thm:firstStep}]
Let us start by noting that
\begin{equation} \label{eq:xyDecomp}
    (\bbx,\bby) \notin \ccalD_{\ccalH} \quad \Rightarrow \quad \bbx - \bby = \sum_{j=1}^{N} \delta_{j} \bbv_{k}
\end{equation}
with at least one nonzero $\delta_{j}$ for $j < K$. Proving that this implies $(\bbx,\bby) \notin \ccalD_{\bbPhi}$ means proving that
\begin{equation} \label{eq:nonzeroEnergyGNN}
    \bbv_{k}^{\Tr} \Big( \sigma \big( \bbH^{f} \bbx \big) - \sigma(\bbH^{f} \bby \big) \Big) \neq 0
\end{equation}
for at least one filter bank $\bbH^{f}$ and one eigenvector $\bbv_{k}$ with $k \leq K$. In fact, \eqref{eq:nonzeroEnergyGNN} means that some of the energy in the difference of the GNN output for $\bbx$ and $\bby$ falls in the low eigenvalues $\lambda_{k} < \lambda_{C}$ guaranteeing that $\bbPhi(\bbx) - \bbPhi(\bby) \notin \Nul(\bbV_{K})$.

Leveraging \eqref{eq:xyDecomp} and the GFT of the filter \eqref{eq:GFToutput}, we can rewrite \eqref{eq:nonzeroEnergyGNN} as
\begin{equation} \label{eq:nonzeroEnergyGNNxy}
    \bbv_{k}^{\Tr} \Big( \sigma \big( \bbH^{f} \bbx \big) - \sigma(\bbH^{f} \bbx - \sum_{j=1}^{N} \delta_{j} h^{f}(\lambda_{j}) \bbv_{j} \big) \Big) \neq 0
\end{equation}
for at least one value of $k \leq K$. Denote $x_{i}^{f} = [\bbH^{f}\bbx]_{i}$ and $\varepsilon_{i}^{f} = \sum_{j=1}^{N} \delta_{j} h^{f}(\lambda_{j}) v_{ij}$ with $[\bbv_{j}]_{i} = [\bbV]_{ij} = v_{ij}$. Then, we can write \eqref{eq:nonzeroEnergyGNNxy} as
\begin{equation} \label{eq:nonzeroEnergyGNNxyElement}
    \sum_{i=1}^{N} v_{ik} \big( \sigma(x_{i}^{f}) - \sigma(x_{i}^{f} - \varepsilon_{i}^{f}) \big) \neq 0
\end{equation}
for at least one value of $k \leq K$. By multiplying and dividing by $\varepsilon_i$ in the summand this is equivalent to
\begin{equation} \label{eq:nonzeroEnergyGNNxyElementBi}
    \sum_{i=1}^{N} v_{ik} \varepsilon_{i}^{f} b_{i}^{f} \neq 0
\end{equation}
where $b_{i}^{f} = (\sigma(x_{i}^{f}) - \sigma(x_{i}^{f} - \varepsilon_{i}^{f}))/\varepsilon_{i}^{f}$ for $\varepsilon_{i}^{f} \neq 0$ and $b_{i}^{f} = \sigma'(x_{i}^{f})$ for $\varepsilon_{i}^{f} = 0$. Note that $b_{i}^{f} > 0$ due to strict monotonicity. Replacing back the value of $\varepsilon_{i}^{f}$ in \eqref{eq:nonzeroEnergyGNNxyElementBi}, we get
\begin{equation} \label{eq:nonzeroEnergyGNNxyElementHj}
    \sum_{i=1}^{N} v_{ik} b_{i}^{f} \sum_{j=1}^{N} \delta_{j} h^{f}(\lambda_{j}) v_{ij} \neq 0
\end{equation}
for at least one value of $k \leq K$. Writing the sum over $i$ as vector multiplication, we get
\begin{equation} \label{eq:nonzeroEnergyGNNxyHj}
    \bbv_{k}^{\Tr} \diag(\bbb^{f}) \sum_{j=1}^{N} \delta_{j} h^{f}(\lambda_{j}) \bbv_{j} \neq 0
\end{equation}
for at least one value of $k \leq K$, and where $\bbb^{f} \in \reals^{N}$ such that $[\bbb^{f}]_{i} = b_{i}^{f}$.

Now, in order to prove this Theorem, \eqref{eq:nonzeroEnergyGNNxyHj} has to hold for at least one value of $k \leq K$. That means that the theorem will be false if and only if
\begin{equation} \label{eq:zeroEnergyGNNxyHj}
\bbV_{K}^{\Tr} \diag(\bbb^{f}) \sum_{j=1}^{N} \delta_{j} h^{f}(\lambda_{j}) \bbv_{j} = \bbzero.
\end{equation}
Let $C_{\sigma}$ be the Lipschitz constant for $\sigma$. Then, multiply \eqref{eq:zeroEnergyGNNxyHj} by $-1$ and add the term $C_{\sigma}\bbV_{K}^{\Tr} \sum_{j=1}^{N} \delta_{j} h^{f}(\lambda_{j})\bbv_{j}$ on both sides to get
\begin{equation} \label{eq:zeroEnergyC}
\bbV_{K}^{\Tr} \big(C_{\sigma} \bbI - \diag(\bbb^{f})\big) \sum_{j=1}^{N} \delta_{j} h^{f}(\lambda_{j}) \bbv_{j} = C_{\sigma}\bbV_{K}^{\Tr}  \sum_{j=1}^{N} \delta_{j} h^{f}(\lambda_{j}) \bbv_{j}.
\end{equation}
So, if \eqref{eq:zeroEnergyC} is true, then \eqref{eq:zeroEnergyGNNxyHj} is true, and thus the Theorem is false. Therefore, we proceed to show that \eqref{eq:zeroEnergyC} cannot hold. To do this, we upper bound the norm of the LHS and show that it is strictly smaller than the norm of the RHS, which would imply that the equality in \eqref{eq:zeroEnergyC} cannot hold. The norm of the LHS can be upper bounded using the submultiplicativity of the operator norm to yield
\begin{equation} \label{eq:LHSbound}
\begin{aligned}
    \Big\| \bbV_{K}^{\Tr} & \big(C_{\sigma} \bbI - \diag(\bbb^{f})\big) \sum_{j=1}^{N} \delta_{j} h^{f}(\lambda_{j}) \bbv_{j} \Big\|^{2} \\
    & \leq \Big\| \bbV_{K}^{\Tr} \big(C_{\sigma} \bbI - \diag(\bbb^{f})\big) \Big\|^{2} \Big\| \sum_{j=1}^{N} \delta_{j} h^{f}(\lambda_{j}) \bbv_{j} \Big\|^{2} \\
    & = \bigg( \max_{i} |C_{\sigma} - b_{i}^{f}|^{2} \bigg) \bigg( \sum_{j=1}^{N} |\delta_{j} h^{f}(\lambda_{j})|^{2} \bigg).
\end{aligned}
\end{equation}
The norm of the RHS of \eqref{eq:zeroEnergyC} becomes
\begin{equation} \label{eq:RHSbound}
    \Big\| C_{\sigma} \bbV_{K}^{\Tr} \sum_{j=1}^{N}\delta_{j} h^{f}(\lambda_{j})  \bbv_{j} \Big\|^{2} = C_{\sigma}^{2} \sum_{j=1}^{K} |\delta_{j} h^{f}(\lambda_{j}) |^{2}
\end{equation}
So we need to prove that
\begin{equation}
    \bigg( \max_{i} |C_{\sigma} - b_{i}^{f}|^{2} \bigg) \sum_{j=1}^{N} |\delta_{j} h^{f}(\lambda_{j})|^{2} < C_{\sigma}^{2} \sum_{j=1}^{K} |\delta_{j} h^{f}(\lambda_{j})  |^{2}
\end{equation}
with strict inequality. Recall that $0 < b_{i}^{f} \leq C_{\sigma}$ for all $i$, and thus $\max_{i} |C_{\sigma} - b_{i}^{f}|^{2} < C_{\sigma}^{2}$ with strict inequality. Thus, since we assume $(\bbx, \bby) \not \in \ccalD_{\ccalH}$ the RHS cannot be zero and it suffices to show
\begin{equation} \label{eq:lowpassIneq}
    \sum_{j=1}^{N} |\delta_{j} h^{f}(\lambda_{j})|^{2} \leq \sum_{j=1}^{K} |\delta_{j}h^{f}(\lambda_{j}) |^{2}.
\end{equation}
Now, \eqref{eq:lowpassIneq} holds under the hypothesis that $h^{f}(\lambda_{j}) = 0$ for $j > K$. Thus, since \eqref{eq:lowpassIneq} is true, then \eqref{eq:zeroEnergyGNNxyHj} is necessarily false, making \eqref{eq:nonzeroEnergyGNNxyHj} true for at least one value of $k$, and thus completing the proof.
\end{proof}

\begin{proof}[Proof of Theorem~\ref{thm:secondStep}]
In this case, let us consider two signals that cannot be discriminated by the graph filter bank
\begin{equation}  \label{eq:xyDecompND}
    (\bbx,\bby) \in \ccalD_{\ccalH} \quad \Rightarrow \quad \bbx - \bby = \sum_{j=K+1}^{N} \delta_{j} \bbv_{k}
\end{equation}
with at least one nonzero $\delta_{j}$. This pair of signals, however, can be discriminated by the GNN as long as $(\bbx,\bby) \notin \ccalD_{\bbPhi}$. This means that
\begin{equation}
    \bbv_{k}^{\Tr} \Big( \sigma(\bbH^{f}\bbx) - \sigma(\bbH^{f}\bby)\Big) \neq 0
\end{equation}
for at least one value of $k \leq K$. Using \eqref{eq:xyDecompND} and denoting by $x_{i}^{f} = [\bbH^{f}\bbx]_{i}$ and $\varepsilon_{i}^{f} = \sum_{j=K+1}^{N} \delta_{j} h^{f}(\lambda_{j})v_{ij}$, we get
\begin{equation}
    \sum_{i=1}^{N} v_{ik} \big( \sigma(x_{i}^{f}) - \sigma(x_{i}^{f}-\varepsilon_{i}^{f})\big) \neq 0
\end{equation}
for at least one value of $k \leq K$. Defining the secant $b_{i}^{f} = (\sigma(x_{i}^{f}) - \sigma(x_{i}^{f} - \varepsilon_{i}^{f}))/\varepsilon_{i}^{f}$ for $\varepsilon_{i}^{f} \neq 0$ and the derivative $b_{i}^{f} = \sigma'(x_{i}^{f})$ for $\varepsilon_{i}^{f} = 0$, recalling that $0<b_{i}^{f}<C_{\sigma}$ and replacing $\varepsilon_{i}^{f}$ by its definition, we get
\begin{equation} \label{eq:nonzeroEnergyGNNxyHjND}
    \sum_{j=K+1}^{N} \delta_{j} h^{f}(\lambda_{j}) \bbv_{k}^{\Tr} \diag(\bbb^{f}) \bbv_{j} \neq 0
\end{equation}
for at least one $k \leq K$.

Note that \eqref{eq:nonzeroEnergyGNNxyHjND} holding for at least one value of $k \leq K$ implies that $(\bbx,\bby) \notin \ccalD_{\bbPhi}$. Now, \eqref{eq:nonzeroEnergyGNNxyHjND} is true for at least one value of $k \leq K$ if and only if, the following statement is false
\begin{equation} \label{eq:zeroEnergyGNNxyHjND}
\sum_{j=K+1}^{N} \delta_{j} h^{f}(\lambda_{j}) \bbV_{K}^{\Tr} \diag(\bbb^{f}) \bbv_{j} = \bbzero.
\end{equation}
From Theorem~\ref{thm:firstStep} we know that $h^{f}(\lambda_{j})=0$ for at least one value of $f \in \{1,\ldots,F\}$. If this is the case for all $f$, then \eqref{eq:zeroEnergyGNNxyHjND} would hold, and the signals will not be discriminable. Thus, we need at least $F=2$ filters, with the second filter such that $h^{f}(\lambda_{j}) \neq 0$ for at least one value of $j > K$. Now, note that, since the filters are integral Lipschitz, they are approximately constant for values of $\lambda > \lambda_{C}$. Thus, let us denote by $h^{f}(\lambda_{j}) = h^{f}$ the constant value of the $f$th filter, for $\lambda > \lambda_{C}$. We further assume that $F \geq 2$, $h^{1} = 0$ and $h^{f} \neq 0$ for at least one value of $f \geq 2$. We can then rewrite \eqref{eq:zeroEnergyGNNxyHjND} as
\begin{equation} \label{eq:zeroEnergyGNNxyHjNDsubspace}
\bbV_{K}^{\Tr} \sum_{j=K+1}^{N} \delta_{j} h^{f} \diag(\bbb^{f}) \bbv_{j} = \bbzero.
\end{equation}
Note that \eqref{eq:zeroEnergyGNNxyHjNDsubspace} holds if and only if the summation yields an element that is in $\Nul(\bbV_{K})$. This means that \eqref{eq:zeroEnergyGNNxyHjNDsubspace} holds if and only if there exists $\bbc = [c_{K+1},\ldots,c_{N}] \in \reals^{N-K}$ such that
\begin{equation} \label{eq:nullSpaceCondition}
\sum_{j=K+1}^{N} \delta_{j} h^{f} \diag(\bbb^{f}) \bbv_{j} = \sum_{j=K+1}^{N} c_{j} \bbv_{j}
\end{equation}
for the given values of $\delta_{j}$ and $\bbb^{f}$ (which depends on the specific pair of signals we are trying to discriminate) and $h^{f}$ (which is designed, or learned from data). We rewrite \eqref{eq:nullSpaceCondition} as
\begin{equation}
\sum_{j=K+1}^{N} \Big( \delta_{j} h^{f} \diag(\bbb^{f}) - c_{j} \bbI \Big) \bbv_{j} = \bbzero
\end{equation}
which has to hold for every one of the $N$ entries of the vector, i.e.
\begin{equation} \label{eq:entrywiseNullSpaceCondition}
    \sum_{j=K+1}^{N} \Big(\delta_{j} h^{f} b_{i}^{f} - c_{j} \Big) v_{ij} = 0
\end{equation}
for all $i \in \{1,\ldots,N\}$. Recall that $\bbV_{N-K} = [\bbv_{K+1},\ldots,\bbv_{N}] \in \reals^{N \times (N-K)}$ and denote by $\bbnu_{i} = [v_{i(K+1)},\ldots,v_{iN}] \in \reals^{N-K}$ the $i$th row of $\bbV_{N-K}$. Note that, since $\rank(\bbV_{N-K}) = N-K$ by definition, then $\mathrm{span}\{\bbnu_{1},\ldots,\bbnu_{N}\} = \reals^{N-K}$. That is, $\{\bbnu_{1},\ldots,\bbnu_{N}\}$ is a set of linearly dependent elements that spans all of $\reals^{N-K}$. With this notation in place, we can write \eqref{eq:entrywiseNullSpaceCondition} as the set of linear equations
\begin{equation}
    (h^{f}b_{i}^{f}\bbdelta - \bbc)^{\Tr} \bbnu_{i} = 0
\end{equation}
for all $i \in \{1,\ldots,N\}$, and where $\bbdelta = [\delta_{K+1},\ldots,\delta_{N}] \in \reals^{N-K}$.

Now, since $\{\bbnu_{1},\ldots,\bbnu_{N}\}$ spans all of $\reals^{N-K}$ and since the inner product of $h^{f}b_{i}^{f}\bbdelta - \bbc$ with every one of the $\bbnu_{i}$ is $0$, then it must hold that
\begin{equation} \label{eq:secondStepCondition}
    \bbc = h^{f}b_{i}^{f} \bbdelta.
\end{equation}
for all $i \in \{1,\ldots,N\}$. In short, for the GNN to fail to discriminate the pair of signals $(\bbx,\bby)$, then \eqref{eq:secondStepCondition} must hold for all $i \in \{1,\ldots,N\}$. For this to happen we need (i) that $h^{f} = 0$ for all $f \in \{1,\ldots,F\}$, a condition that was already ruled out by hypothesis; or (ii) that $\bbdelta = \bbzero$, which is also ruled out by hypothesis, i.e at least one value $\delta_{j}$ is nonzero $(\bbx,\bby) \in \ccalD_{\ccalH}$ to hold; or (iii) to have $b_{i}^{f} = b^{f}$ for all $i$, i.e. to have a pair of signals $(\bbx,\bby)$ such that their secants $b_{i}^{f}$ are constant for all values of $i$. This completes the proof.
\end{proof}

\begin{proof}[Proof of Corollary \ref{cor:setEquivalence}]
From the proof of Theorem~\ref{thm:secondStep} we know that the pair of signals $(\bbx,\bby) \in \ccalD_{\ccalH}$ are discriminable as long as \eqref{eq:nonzeroEnergyGNNxyHjND} holds for at least one $k \leq K$. This is the case if and only if \eqref{eq:zeroEnergyGNNxyHjND} is false. Thus, if \eqref{eq:zeroEnergyGNNxyHjND} is true, then the signals are not discriminable.

Note that by setting $h^{f}(\lambda_{j})=0$ for $\lambda>\lambda_{C}$ for all $f$, then \eqref{eq:zeroEnergyGNNxyHjND} is true for all pairs of signals $(\bbx,\bby) \in \ccalD_{\ccalH}$. This implies that every element in $(\bbx,\bby) \in \ccalD_{\ccalH}$ also satisfies $(\bbx,\bby) \in \ccalD_{\bbPhi}$, so that $\ccalD_{\ccalH} \subseteq \ccalD_{\bbPhi}$.

Since, from Theorem~\ref{thm:firstStep} we have $\ccalD_{\bbPhi} \subseteq \ccalD_{\ccalH}$, we can conclude that $\ccalD_{\ccalH} \equiv \ccalD_{\bbPhi}$, completing the proof.
\end{proof}

\begin{proof}[Proof of Corollary \ref{cor:tanhSet}]
From Theorem~\ref{thm:secondStep} we know that the pair of signals $(\bbx,\bby) \in \ccalD_{\ccalH}$ will be nondiscriminable as long as $b_{i}^{f}$ is a constant over $i$ for all filters $f$. If we have $\sigma = \tanh$, then
\begin{equation}
    b_{i}^{f} = \frac{\tanh(x_{i}^{f}) - \tanh(x_{i}^{f} - \varepsilon_{i}^{f})}{\varepsilon_{i}^{f}}.
\end{equation}
for $\varepsilon_{i}^{f} = \sum_{j=K+1}^{N} \delta_{j} v_{ij}$. Accounting for the $\varepsilon_i^f=0$ case is inconsequential as will become apparent later in the proof. Let us consider the signal pairs that will not be discriminated. Those are the ones that satisfy
\begin{equation} \label{eq:noClosedFormSol}
    \tanh(x_{i}^{f}) = b^{f} \varepsilon_{i}^{f} + \tanh(x_{i}^{f}-\varepsilon_{i}^{f})
\end{equation}
for all $i \in \{1,\ldots,N\}$, and where $0 < b^{f} < 1$ is the value of the secant shared across all nodes. Given an arbitrary output of the filter phase at node $i$, $x_{i}^{f}$, then the value of $\varepsilon_{i}^{f}$ has to satisfy \eqref{eq:noClosedFormSol}. Let us denote this value by $\varepsilon_{i}^{f\ast}(b^{f})$. Note that, for each, node, the value of $\varepsilon_{i}^{f\ast}$ is different, but all depend on the same value of $b^{f}$. Let us collect those values in the vector $\bbvarepsilon^{f\ast}(b^{f}) \in \reals^{N}$.

Now, for the secant to be the same at all nodes, then the values of $\bbdelta^{f} = [\delta_{K+1}^{f},\ldots,\delta_{N}^{f}] \in \reals^{N-K}$ has to satisfy
\begin{equation} \label{eq:overdeterminedSystem}
    \bbV_{N-K}^{\Tr} \bbdelta^{f} = \bbvarepsilon^{f\ast}(b^{f}).
\end{equation}
This is an overdetermined system of $N$ linear equations with $N-K$ unknowns. We also know that $\bbV_{N-K}$ has rank $N-K$, which implies that \eqref{eq:overdeterminedSystem} has either a unique solution, or no solution at all. This means that, for a fixed $\bbx$ and any particular value of $b^{f}$, there is at most a single signal $\bby$ with a constant secant at all nodes equal to $b^{f}$. This means that the space of signal pairs that yield a constant secant is, at most, one-dimensional. Therefore, since $\ccalD_{\ccalH}$ has dimension $N-K$, we conclude that there are signals in $\ccalD_{\ccalH}$ that will be discriminated by the GNN, as long as $N-K > 1$.
\end{proof}

\end{document}